\documentclass{article}
\usepackage{spconf,amsmath,graphicx}

\usepackage[ruled]{algorithm2e}
\usepackage{booktabs}
\usepackage{multirow}
\usepackage{tabularx}
\usepackage{graphicx}
\usepackage{amsfonts}
\usepackage{threeparttable}
\usepackage{makecell}

\usepackage{lipsum}

\newcommand\blfootnote[1]{%
  \begingroup
  \renewcommand\thefootnote{}\footnotetext{#1}%
  \addtocounter{footnote}{-1}%
  \endgroup
}

\title{Learning Music Sequence Representation from Text Supervision}
\name{Tianyu Chen\textsuperscript{\rm 1} \textsuperscript{\rm 3} $^{\dagger}$, Yuan Xie\textsuperscript{\rm 4} $^{\dagger}$  , Shuai Zhang \textsuperscript{\rm 1} \textsuperscript{\rm 3}, Shaohan Huang \textsuperscript{\rm 2},  Haoyi Zhou \textsuperscript{\rm 1} \textsuperscript{\rm 3}, Jianxin Li \textsuperscript{\rm 1} \textsuperscript{\rm 3} }
\address{
\textsuperscript{\rm 1}BDBC, Beihang University, China\  \textsuperscript{\rm 2}Microsoft Research Asia, China \\
\textsuperscript{\rm 3}SKLSDE, Beihang University \
\textsuperscript{\rm 4}The Institute of Acoustics of the Chinese Academy of Sciences, China 
}

\begin{document}
%
\maketitle
\begin{abstract}



Music representation learning is notoriously difficult for its complex human-related concepts contained in the sequence of numerical signals. To excavate better \textbf{MU}sic \textbf{SE}quence \textbf{R}epresentation from labeled audio, we propose a novel text-supervision pre-training method, namely \textbf{MUSER}. MUSER adopts an audio-spectrum-text tri-modal contrastive learning framework, where the text input could be any form of meta-data with the help of text templates while the spectrum is derived from an audio sequence. Our experiments reveal that MUSER could be more flexibly adapted to downstream tasks compared with the current data-hungry pre-training method, and it only requires 0.056\% of pre-training data to achieve the state-of-the-art performance. \blfootnote{$^{\dagger}$ indicates equal contribution. The corresponding author is Jianxin Li. This work is supported by the NSFC through grant No.61872022. }


\end{abstract}
\section{Introduction}
\label{sec:intro}



Music is an inseparable part of human culture, containing complex emotional or narrative concepts in audio sequences. It evokes a crucial question for computer scientists - \textit{``Can our computer understand the meaning of music?''} Over the recent years, various music understanding benchmarks have been built, including music tagging~\cite{Choi2016AutomaticTU, Pons2018EndtoendLF, Won2019TowardIM} and genre classification~\cite{Bahuleyan2018MusicGC}. 

Inspired by the success of deep learning on computer vision and natural language processing, researchers introduce large-scale pre-training techniques into music sequence representation learning~\cite{Cramer2019LookLA, Pons2019musicnnPC}. The learned representation can be further transferred to different downstream tasks, achieving better performance than former end-to-end methods. However, pre-training is data-hungry. Further improvements on downstream tasks rely heavily on more hand-crafted tags~\cite{Oord2014TransferLB} or other metadata~\cite{Lee2019RepresentationLO, Huang2020LargeScaleWC}, where the music labels require professional knowledge, becoming luxury.

\begin{figure}
    \centering
    \includegraphics[width=0.85\linewidth]{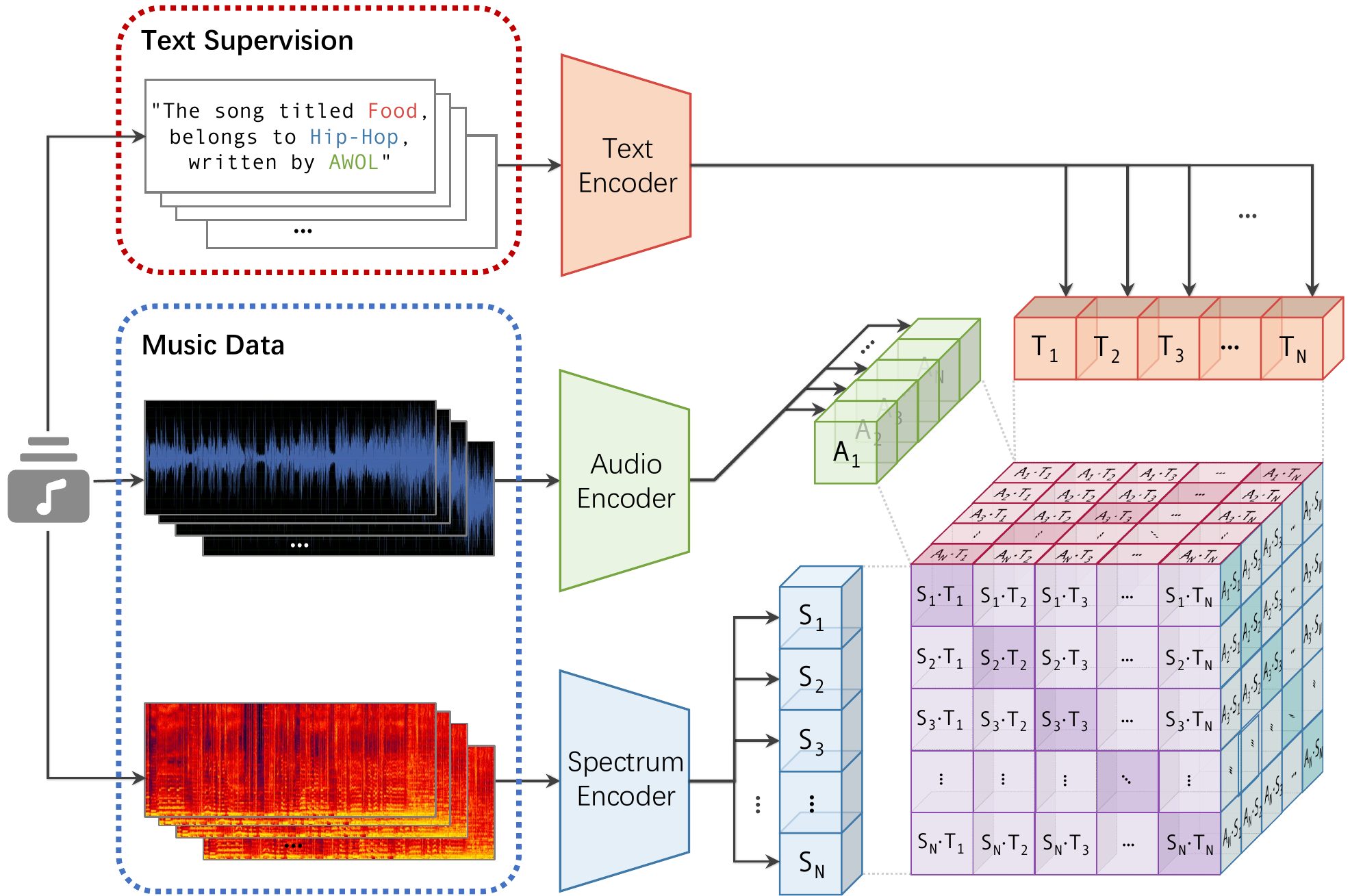}
    \caption{The framework of MUSER.}
   \label{fig:framework}
  \vspace{-4mm}
\end{figure}

A question is naturally brought, \textit{``Are existing labels enough to learn better music representation?''} Music data is relatively quantity-small but with sufficient supervision information in their text-form metadata (e.g., lyrics, album descriptions, lyricist, composer, singer, comments), which are still under-explored. In this paper, we propose a novel text-supervision method to learn directly from text-form metadata, called MUSER. First, we convert different forms of metadata into a unified plain text format (e.g., one typical text format contains song name and genre class - ``The song titled \texttt{Food}, belongs to \texttt{Hip-Hop}'' ). Then, the plain text sequence and the audio sequence of music will be encoded respectively into a shared embedding space. Also, the music spectrum is encoded to learn text-form concepts from multi-view. By using the shared embedding space, we can apply CLIP-style~\cite{radford2021learning}\cite{Ma2021ASL} contrastive learning, which aims to distinguish music sequence by their corresponding text. After contrastive pre-training stage on limited training data, we fine-tune the encoders on genre classification and auto-tagging benchmarks, achieving comparable or even better results than state-of-the-art pre-training methods. Our contributions are:
(1) We are the first to introduce text supervision for exploring the fine-grained feature of distributed songs;
(2) Additional spectrogram encoder that greatly improves data efficiency of the CLIP-style framework;
(3) A novel tri-modal contrastive pre-training framework - MUSER and new state-of-the-art on music-related benchmarks.

\vspace{-2mm}
\section{Method}
\vspace{-2mm}
\subsection{The Framework of MUSER}

As illustrated in Fig.\ref{fig:framework}, three individual encoders are involved in our MUSER framework. To fully excavate the text supervision signals, we initialize our text encoder and spectrum encoder from OpenAI pre-trained CLIP~\cite{radford2021learning}, where the music spectrum encoder is considered as a special ``image encoder". To get more information, we use templates to stitch together various music metadata. Templates splice descriptive phrases into a complete sentence as text input. And a befitting template could help define the decision boundaries for classification tasks.



\textbf{Text Encoder}: The text encoder is a base-size Transformer~\cite{vaswani2017attention} network. Before operating on text tokens, the encoder employs a byte pair encoding (BPE) tokenizer to convert plain text into a sequence of discrete tokens. The text sequence will be bracketed with special tokens, \texttt{[SOS]} and  \texttt{[EOS]} . The activations of the \texttt{[EOS]} at the highest layer will be treated as the feature representation of the text and then linearly projected into the shared embedding space.

\textbf{Music Spectrum Encoder}: We use a ResNet-50 image encoder as music spectrum encoder~\cite{he2016deep}. Unlike the raw version, we change the global average pooling layer into an attention pooling mechanism, where a ``transformer-style'' $QKV$ attention\cite{Wang2018NonlocalNN} is implemented. The image representation output from the ResNet encoder will serve as a query.

\textbf{Music Audio Encoder}:  ESResNeXt~\cite{guzhov2021esresne} serves as our audio encoder, which is  used in AudioClip \cite{guzhov2021audioclip}. Based on ResNet-50,  ESResNext includes a trainable time-frequency transformation with complex frequency B-spline wavelets~\cite{Teolis1998ComputationalSP}.



\vspace{-2mm}
\subsection{Tri-modal Contrastive Learning}

In this section,  we will introduce our training algorithm in detail. Given a batch of labeled input music sequences, we denote the $i$-th music audio sequence as $A_i$, the corresponding text label as $T_i$. Then a music spectrum $S_i$ could be calculated by Short-Time Fourier Transform (STFT).


Then we can get the embeddings of the three modals by forwarding the input into their encoders:
\begin{equation}
    E_{A_i},  E_{T_i}, E_{S_i} = F_\text{aud}(A_i),F_\text{txt}(T_i), F_\text{spec}(S_i) .
\end{equation}

The core idea of contrastive learning is to pull the similar embeddings closer while pushing the other embeddings away. In our scenario, we pull the $E_{T_i}$ closer to $E_{A_i}$ and $E_{S_i}$ while push the $E_{T_i}$ far away from $E_{A_j}$ and $E_{S_j}$, where $j$-th example is another music sequence in the same batch. Now we get the optimization goal for modal $Q$ and $K$:

\begin{equation}
    L(i, j; \theta_Q, \theta_K )_ = -\log \frac{\exp(D(E_{Q_i}, E_{K_i}) / \tau)}{ \sum^{N}_{j=1, j \neq i} \exp(D(E_{Q_i}, E_{K_j}) / \tau)},
\end{equation}
where $N$ is the batch size and $\tau$ is a learnable parameter. $\theta_Q$ is the parameter of the encoder for modal $Q$ and $\theta_K$ is the parameter of the encoder for another modal $K$. $D$ is a distance function to evaluate the similarity between $E_{Q_i}$ and $E_{K_i}$. As depicted in Fig.\ref{fig:framework} , for all three modals, we optimize them together with asymmetric contrastive loss:
\begin{equation}
    \mathcal{L}(i, j; \theta_A, \theta_T, \theta_S) = \mathcal{L}(i, j; \theta_A, \theta_T) + \mathcal{L}(i, j; \theta_S, \theta_T),
\end{equation}
where $\theta_A, \theta_T, \theta_S$ represent the parameters of audio encoder, text encoder and music spectrum encoder, respectively. A more detailed training algorithm is illustrated in Alg. \ref{alg:learning}.





    
    


\setlength{\algomargin}{1em}
\begin{algorithm}[t]
\caption{Contrastive Learning of MUSER.}
\label{alg:learning}
\LinesNumbered
\SetKwFor{ForAll}{for all}{do}{end}
\KwData{all labeled training music sequence $\mathcal{A}$, text $\mathcal{T}$, spectrum $\mathcal{S}$ pairs, and label $\mathcal{Y}$.}
\KwIn{encoders $F_\text{aud}$, $F_\text{txt}$, $F_\text{spec}$; weights $W_a$, $W_t$, $W_s$; temperature parameter $\tau$; batch size $n$.}
\While{not done}{
    Sample batches $(A_i, T_i, S_i, Y_i) \sim (\mathcal{A, T, S, Y})$.\\
    \ForAll{$(A_i, T_i, S_i, Y_i)$}{
        $E_{A_i}, E_{T_i}, E_{S_i} \leftarrow F_\text{aud}(A_i) , F_\text{txt}(T_i) , F_\text{spec}(S_i)$. \\
        $E_{A_i}, E_{T_i}, E_{S_i} \leftarrow W_a E_{A_i},  W_t E_{T_i}, W_s E_{S_i}$.\\
        Compute logits $\widehat{Y}_{AT}, \widehat{Y}_{TA}, \widehat{Y}_{ST}, \widehat{Y}_{TS}$ as: e.g., $\widehat{Y}_{AT_i} = E_{A_i}  E_{T_i}  e^ \tau$. \
        Compute losses $\ell_{AT_i}, \ell_{TA_i}, \ell_{ST_i}, \ell_{TS_i}$ as: e.g., $\ell_{AT_i} = \text{CrossEntropy}(\widehat{Y}_{AT_i}, Y_i, axis=0)$.\\
        Compute overall loss $\ell_i = (\ell_{AT_i} + \ell_{TA_i} + \ell_{ST_i} + \ell_{TS_i})/4$.\\
    }
    Update encoders and weights with loss $\mathcal{L} = \sum_i{\ell_i}.$
}
\end{algorithm}


\section{Experiment Setup}

We experiment with different music benchmarks with an extra small music dataset for pre-training. For fair comparisons, we include state-of-the-art methods as baselines and report results under the same evaluation settings.   

\subsection{Benchmark Tasks}

Two downstream music-related tasks could benchmark the performance of music sequence representation learning: (1) genre classification (2) automatic tagging.


\subsubsection{Genre Classification}

Genre classification involves assigning the most appropriate
genre from a given song. We choose GTZAN as the dataset for this task, which contains 1,000 tracks of 30-second length. In consideration of several annotation errors on the dataset~\cite{Sturm2013TheGD}, we adopt the ``fault-filtered” split~\cite{Kereliuk2015DeepLA} to minimize the impact of error labels. The filtered training set (443 tracks) is regarded as a part of the fusion training set. We use accuracy as the evaluation metrics.


\subsubsection{Automatic Tagging}
Automatic tagging aims to find the most appropriate tags for target songs, also viewed as a multi-label classification problem. We choose MagnaTagATune (MTT)~\cite{Law2009EvaluationOA} as the benchmark dataset  To ensure enough training data for each tag, we limit the vocabulary to the top 50 most popular tags. We use the standard 12:1:3 train/val/test split for MTT~\cite{Oord2014TransferLB}. Two macro-averaged over tags metrics are reported: area under the receiver operating characteristic curve (ROC-AUC), and average precision (AP).

\subsection{Pre-training Datasets}

Free Music Archive (FMA): FMA is a large-scale dataset for evaluating several tasks in Music Information Retrieval. We use the small subset of FMA, a balanced subset containing 8,000 clips. We rely on templates to concatenate the genre, parent genre, and top-level tag of each audio together.

Traditional models can only be trained on a single dataset, which tends to result in poor generalization performance. In contrast, the powerful transformer-based text encoder enables MUSER to fuse all kinds of datasets with different annotations together. We fuse two benchmark datasets with an extra FMA dataset for pre-training.

In total, we use around 23,500 clips as pre-training data, only 0.056\%  of the data for the state-of-the-art pre-training method. The details are depicted in Table \ref{tab:data.stats}.

\begin{table}[t]
  \centering
  \caption{Datasets for Music Sequence Pre-training.}
    \begin{tabular}{lll}
    \toprule
    Method & Source & Audio / Text \\
    \midrule
    VGGish\cite{hershey2017cnn} & YouTube-8M & 350000h / 8M \\
    CLMR\cite{Spijkervet2021ContrastiveLO} & Not mentioned & 2200h / 260k \\
    CALM\cite{Castellon2021CodifiedAL} & Jukebox & 240000h / 1.2M \\
    \hline
    \multirow{4}{*}{\textbf{MUSER}} & FMA (small subset) & 66.7h / 8k \\
          & MTT (train) & 127h / 15k \\
          & GTZAN (train) & 3.7h / 0.4k \\
          & \textbf{Total} & \textbf{195.8h / 23.5k} \\
    \bottomrule
    \end{tabular}%
  \label{tab:data.stats}%
\end{table}%

\subsection{Baseline Methods}

A handful of recent music sequence representation learning methods will be included in our comparison:

\textbf{VGGish~\cite{hershey2017cnn}}: Pre-trained on a large-scale video dataset (AudioSet \cite{gemmeke2017audio}) with a classification task, VGGish is the first to explore the best performance of fully connected DNNs using CNN Architectures.

\textbf{CLMR~\cite{Spijkervet2021ContrastiveLO}}: CLMR also noticed the problem of expensive music labels. It is the first work to introduce the contrastive pre-training techniques, which enable unsupervised music sequence representation learning.

\textbf{CALM~\cite{Castellon2021CodifiedAL}}: CALM is first proposed for unconditional speech generation. It codifies a high-rate continuous audio sequence into low-rate discrete codes. Then a language model is trained on resulting codified audio and optional meta-data to produce high-quality contextual representations.

\subsection{Text Template Design}
The core idea of our approach is to learn musical perception from text supervision formed by manual templates. Traditional classification methods map their metadata into a numeric \texttt{id} of label (e.g, \texttt{0} for ``jazz" and \texttt{1} for ``hip-hop"). This leads to inconsistency between different music datasets where similar tags can be represented by different numerical \texttt{id}s. Instead of label mapping, text input takes advantage of semantic similarity, guiding a fair distribution in embedding space.

 Utilizing manually designed templates to combine music metadata into unified text input, MUSER could bridge the distribution gap. The template helps to restrict the decision boundary, making the embedding space more targeted. Finding an appropriate template is an art - requiring both domain expertise and an understanding of the text encoder's inner workings. We find the text template \texttt{"a song of \{genre\}, belongs to \{tag\}"}. to be a good default that helps clarify the concepts in music metadata. 

During testing, the similarity of music sequence with each class-filled text template will be calculated to find the most similar class.

\begin{table}[t]
  \centering
  \caption{Performance on music understanding benchmarks.}
  \resizebox{\linewidth}{!}{
  \begin{threeparttable}
    \begin{tabular}{lccc}
    \toprule
    Method & Tags (AUC) & Tags (AP) & Genere (ACC) \\
    \midrule
    VGGish  & 89.4 & 42.2 & 75.2 \\
    CLMR  & 89.4  & 36.1  & 68.6 \\
    CALM  & 91.5  & 41.4  & 79.7 \\
    AE only (MT., PT) & 88.7 & 38.4 & 59.7 \\
    AE only (MT., PT+FT) & 88.9 & 38.9 & 76.9 \\
    State-of-the-art   & \textbf{91.5}   & 42.2  & 82.1 \\
    \hline
    MUSER (AE only) & 87.5  & 36.3 & 66.6 \\
    MUSER (w/o spec) & 88.1  & 39.6 & 75.2 \\
    MUSER (PT) & 88.7  & 41.6  & 72.6 \\
    \textbf{MUSER (PT+FT) } & 89.5 & \textbf{43.0}  & \textbf{82.5} \\
    \bottomrule
    \end{tabular}
    \begin{tablenotes}
        \footnotesize
        \item[1] MT. refers to Multi-task.
        \item[2] AE only refers to only with Audio-Encoder.
    \end{tablenotes}
    \end{threeparttable}}
  \label{tab:res.all}%
\end{table}%

\section{Results And Discussion}

We first show the results of MUSER, compared with baseline pre-training methods. Then we investigate the text semantics learned by MUSER and provide suggestions for text template engineering. Finally, we study the high data efficiency of MUSER under few-shot settings.

\subsection{Role of Text Encoder}
Three settings are proposed to prove the text-supervision effect: 
(1) \textbf{MUSER (AE only)}: Without text-supervision, we convert the metadata into numerical labels and utilize the audio encoder for training and test.
(2) \textbf{MUSER (PT)}: Pre-train MUSER encoders on the fused dataset, then directly test them on benchmark tasks.
(3) \textbf{MUSER (PT+FT)}: After text-supervision pre-training, we continual fine-tune all encoders on benchmark dataset train set.


In Table \ref{tab:res.all}, we observe that our MUSER (PT+FT) outperforms the state-of-the-art\footnote{We only compared with pre-training models that use 30-second clips as training data for fair.} on both tagging and genre classification tasks. It is not surprising that without text supervision, only fine-tuning the audio encoder with numerical labels leading to poor performance. And a small portion of music domain data with MUSER pre-training can provide a good initialization for further transfer. The fine-tuning strategy also works in our contrastive setting and benefits a better result.

\begin{figure}[t]
    \centering
    \includegraphics[width=\linewidth]{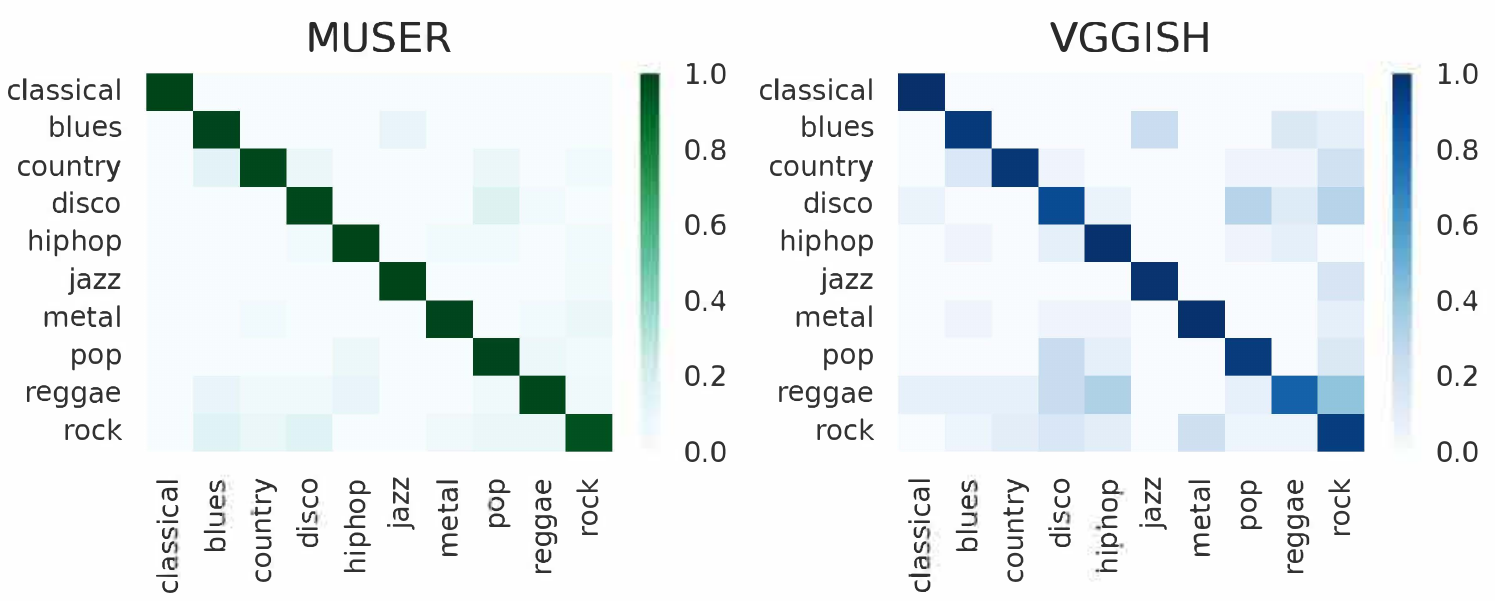}
    \caption{Confusion matrices on GTZAN genre classification.}
    \label{fig:res.confuse}
\end{figure}

From Fig.\ref{fig:res.confuse}, directly learning from plain text, our MUSER can more easily distinguish similar genres like ``rock'' and ``reggae'', where VGGish gets confused.
 
In Table \ref{tab:template}, we analyze the influence of different text templates on MTAT. The selection of text templates is a key factor to our MUSER. Of several text templates we tried, ``a song of \{genre\}, belongs to \{tag\}, whose style is \{style\}'' performs best. We assume the optimal template could be found by reinforcement learning , left for a further study.  

\begin{table}[t]
  \centering
  \caption{Performance of different text templates on MTAT.}
  \resizebox{\linewidth}{!}{
    \begin{tabular}{lcc}
    \toprule
    Template & Tags (AUC) & Tags (AP) \\
    \midrule
    No template & 88.5 & 41.9 \\
    \midrule
    \makecell[l]{``tags for the \{genre\}  music is \{tag\}''}  & 88.3 & 41.4 \\
    \midrule
    \makecell[l]{``the  \{genre\} music is characterized \\ by \{tag\}'' }  & 88.5 & 42.0 \\
    \midrule
    \makecell[l]{``a song of \{genre\}, belongs to \{tag\}, \\whose style is \{style\}''} & 89.5 & 43.0 \\
    \bottomrule
    \end{tabular}%
    }
  \label{tab:template}%
\end{table}%


\vspace{-3mm}
\subsection{Role of Spectrum Encoder}

We observe that the music spectrum encoder benefits contrastive pre-training. Generated by analyzing the music signal, the spectrum does not introduce additional information but enables the MUSER encoders to learn music from different views. From Table \ref{tab:res.all}, MUSER(without spec) settings remove spectrum encoder during pre-training and fine-tuning, we find the spectrum brings a 7.3\% improvement on genre accuracy and 3.4\% improvement on tagging average precision. 



\begin{figure}
    \centering
    \includegraphics[width=\linewidth]{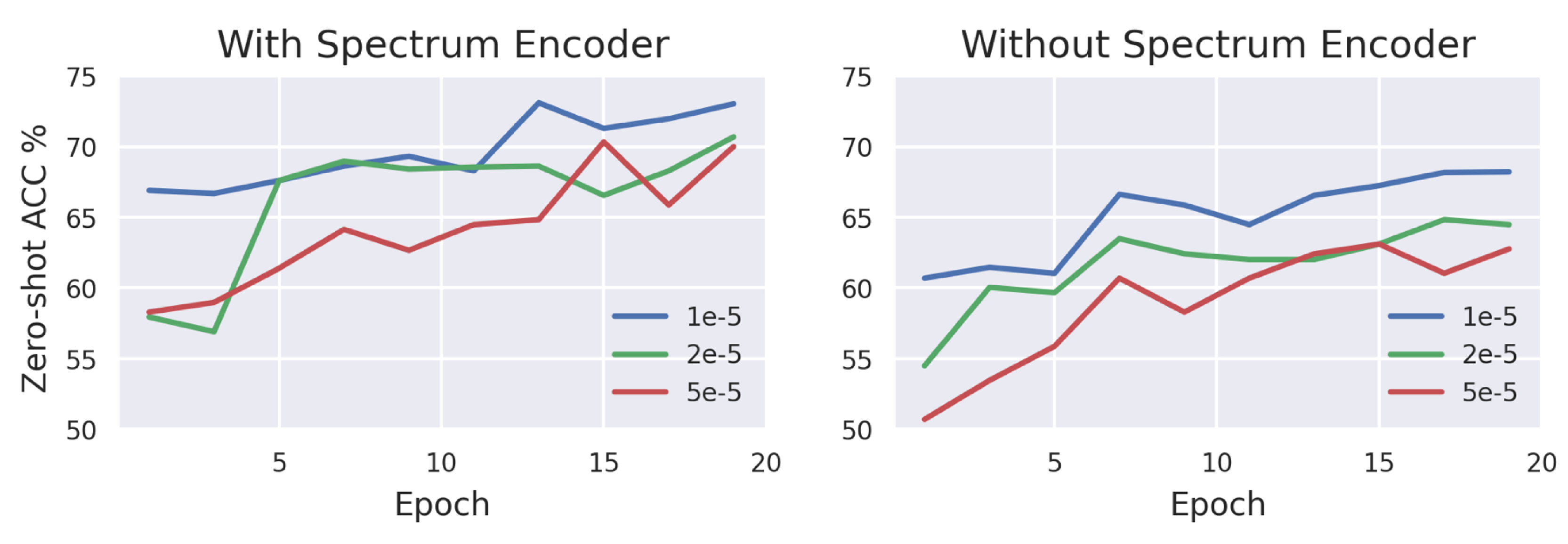}
    \caption{Ablation experiments w/o spectrum encoder. Evaluate with zero-shot accuracy on GTZAN for the first 20 epochs. }
    \label{fig:res.LR}
\end{figure}

\vspace{-2mm}
\subsection{Data Efficiency}

\begin{figure}
    \centering
    \includegraphics[width=\linewidth]{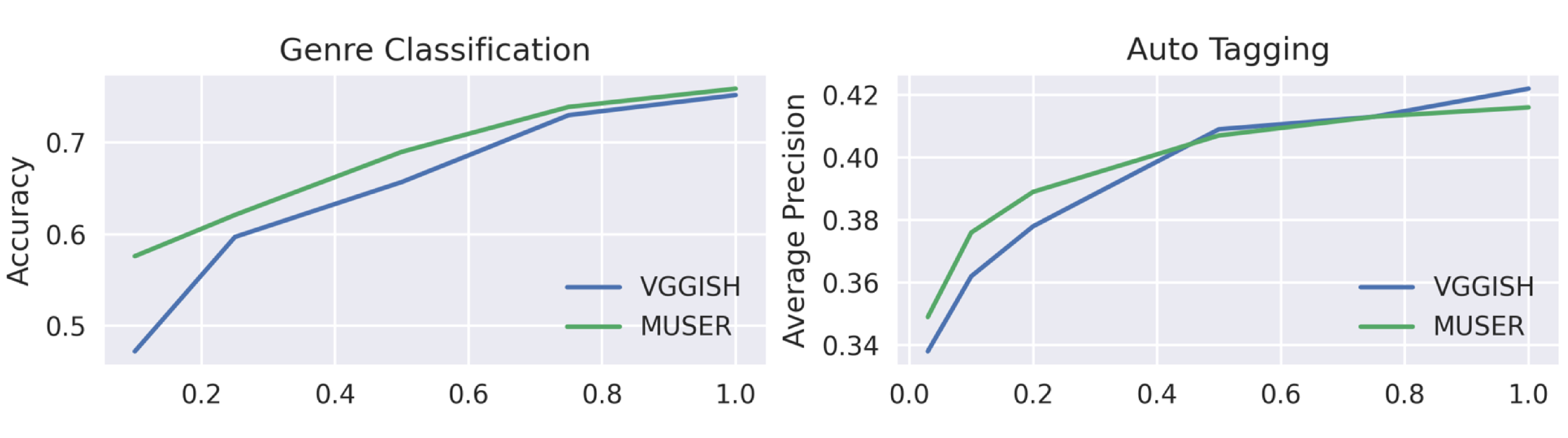}
    \caption{Comparisons of fine-tuning efficiency on \% ratio of training samples for different downstream tasks, with only FMA-small dataset for pre-training.}
    \label{fig:res.few}
\end{figure}

In Fig.\ref{fig:res.few}, we illustrate the performance of our method and VGGish with limited training examples for downstream tasks. The quantity of pre-training data used is less than 0.1\% of other pre-trained models. By less than 40\% of training data, our pre-training method can exhibit a better few-shot ability than VGGish. We hypothesize traditional pre-trained methods need a new task-related network layer on top of them, which is hard to  well-initialize with limited data. 

\vspace{-2mm}
\section{Conclusions}


In this paper, we explore the possibility of learning music sequence representation from text supervision with contrastive pre-training. Compared with SOTA pre-training methods, our MUSER can perform competitively on downstream tasks with far less labeled data. We also reveal that continual fine-tuning with task-related data can further improve performance. For future work, we are interested in exploring the performance boundary of our MUSER method with a large-scale music dataset. Furthermore, prompt-based fine-tuning is promising where the encoders remain fixed. A proper text template can contribute to SOTA performance. This technique may break the boundary between music sequence and natural language description, which helps flexibly generalize pre-trained models to data-limited downstream tasks.



\newpage

\bibliographystyle{IEEEbib}
\bibliography{strings,refs}

\begin{thebibliography}{10}

\bibitem{Choi2016AutomaticTU}
Keunwoo Choi, Gy{\"o}rgy Fazekas, and M.~Sandler,
\newblock ``Automatic tagging using deep convolutional neural networks,''
\newblock in {\em ISMIR}, 2016.

\bibitem{Pons2018EndtoendLF}
Jordi Pons, Oriol Nieto, Matthew Prockup, Erik~M. Schmidt, A.~F. Ehmann, and
  X.~Serra,
\newblock ``End-to-end learning for music audio tagging at scale,''
\newblock in {\em ISMIR}, 2018.

\bibitem{Won2019TowardIM}
Minz Won, Sanghyuk Chun, and X.~Serra,
\newblock ``Toward interpretable music tagging with self-attention,''
\newblock {\em ArXiv}, vol. abs/1906.04972, 2019.

\bibitem{Bahuleyan2018MusicGC}
Hareesh Bahuleyan,
\newblock ``Music genre classification using machine learning techniques,''
\newblock {\em ArXiv}, vol. abs/1804.01149, 2018.

\bibitem{Cramer2019LookLA}
J.~Cramer, Ho-Hsiang Wu, J.~Salamon, and J.~Bello,
\newblock ``Look, listen, and learn more: Design choices for deep audio
  embeddings,''
\newblock in {\em ICASSP}, 2019.

\bibitem{Pons2019musicnnPC}
Jordi Pons and X.~Serra,
\newblock ``musicnn: Pre-trained convolutional neural networks for music audio
  tagging,''
\newblock in {\em ISMIR}, 2019.

\bibitem{Oord2014TransferLB}
A{\"a}ron van~den Oord, S.~Dieleman, and B.~Schrauwen,
\newblock ``Transfer learning by supervised pre-training for audio-based music
  classification,''
\newblock in {\em ISMIR}, 2014.

\bibitem{Lee2019RepresentationLO}
Jongpil Lee, Jiyoung Park, and Juhan Nam,
\newblock ``Representation learning of music using artist, album, and track
  information,''
\newblock in {\em ICML}, 2019.

\bibitem{Huang2020LargeScaleWC}
Qingqing Huang, A.~Jansen, Li~Zhang, D.~Ellis, R.~Saurous, and John~R.
  Anderson,
\newblock ``Large-scale weakly-supervised content embeddings for music
  recommendation and tagging,''
\newblock in {\em ICASSP}, 2020.

\bibitem{radford2021learning}
Alec Radford, Jong~Wook Kim, Chris Hallacy, A.~Ramesh, Gabriel Goh, Sandhini
  Agarwal, Girish Sastry, Amanda Askell, Pamela Mishkin, Jack Clark, Gretchen
  Krueger, and Ilya Sutskever,
\newblock ``Learning transferable visual models from natural language
  supervision,''
\newblock in {\em ICML}, 2021.

\bibitem{Ma2021ASL}
Teli Ma, Shijie Geng, Mengmeng Wang, Jing Shao, Jiasen Lu, Hongsheng Li, Peng
  Gao, and Yu~Qiao,
\newblock ``A simple long-tailed recognition baseline via vision-language
  model,''
\newblock {\em ArXiv}, vol. abs/2111.14745, 2021.

\bibitem{vaswani2017attention}
Ashish Vaswani, Noam Shazeer, Niki Parmar, Jakob Uszkoreit, Llion Jones,
  Aidan~N Gomez, {\L}ukasz Kaiser, and Illia Polosukhin,
\newblock ``Attention is all you need,''
\newblock in {\em NIPS}, 2017.

\bibitem{he2016deep}
Kaiming He, Xiangyu Zhang, Shaoqing Ren, and Jian Sun,
\newblock ``Deep residual learning for image recognition,''
\newblock in {\em CVPR}, 2016.

\bibitem{Wang2018NonlocalNN}
X.~Wang, Ross~B. Girshick, Abhinav~Kumar Gupta, and Kaiming He,
\newblock ``Non-local neural networks,''
\newblock in {\em CVPR}, 2018.

\bibitem{guzhov2021esresne}
Andrey Guzhov, Federico Raue, J{\"o}rn Hees, and Andreas Dengel,
\newblock ``Esresne (x) t-fbsp: Learning robust time-frequency transformation
  of audio,''
\newblock in {\em IJCNN}, 2021.

\bibitem{guzhov2021audioclip}
Andrey Guzhov, Federico Raue, J{\"o}rn Hees, and Andreas Dengel,
\newblock ``Audioclip: Extending clip to image, text and audio,''
\newblock {\em arXiv preprint arXiv:2106.13043}, 2021.

\bibitem{Teolis1998ComputationalSP}
Anthonio Teolis,
\newblock ``Computational signal processing with wavelets,''
\newblock in {\em Applied and numerical harmonic analysis}, 1998.

\bibitem{Sturm2013TheGD}
Bob~L. Sturm,
\newblock ``The gtzan dataset: Its contents, its faults, their effects on
  evaluation, and its future use,''
\newblock {\em ArXiv}, vol. abs/1306.1461, 2013.

\bibitem{Kereliuk2015DeepLA}
Corey Kereliuk, Bob~L. Sturm, and J.~Larsen,
\newblock ``Deep learning and music adversaries,''
\newblock {\em IEEE Trans Multimedia}, 2015.

\bibitem{Law2009EvaluationOA}
Edith Law, Kris West, Michael~I. Mandel, Mert Bay, and J.~S. Downie,
\newblock ``Evaluation of algorithms using games: The case of music tagging,''
\newblock in {\em ISMIR}, 2009.

\bibitem{hershey2017cnn}
Shawn Hershey, Sourish Chaudhuri, Daniel~PW Ellis, Jort~F Gemmeke, Aren Jansen,
  R~Channing Moore, Manoj Plakal, Devin Platt, Rif~A Saurous, Bryan Seybold,
  et~al.,
\newblock ``Cnn architectures for large-scale audio classification,''
\newblock in {\em ICASSP}, 2017.

\bibitem{Spijkervet2021ContrastiveLO}
Janne Spijkervet and J.~Burgoyne,
\newblock ``Contrastive learning of musical representations,''
\newblock in {\em ISMIR}, 2021.

\bibitem{Castellon2021CodifiedAL}
Rodrigo Castellon, Chris Donahue, and Percy Liang,
\newblock ``Codified audio language modeling learns useful representations for
  music information retrieval,''
\newblock in {\em ISMIR}, 2021.

\bibitem{gemmeke2017audio}
Jort~F Gemmeke, Daniel~PW Ellis, Dylan Freedman, Aren Jansen, Wade Lawrence,
  R~Channing Moore, Manoj Plakal, and Marvin Ritter,
\newblock ``Audio set: An ontology and human-labeled dataset for audio
  events,''
\newblock in {\em ICASSP}, 2017.

\end{thebibliography}

\end{document}